\documentclass[preprint,12pt]{elsarticle}
\usepackage{amssymb}
\usepackage{amsmath}
\usepackage{color}
\usepackage{psfrag}
\usepackage{commath}

\journal{NDT \& E International}
\begin{document}
\begin{frontmatter}
\title{Influence of specimen velocity on the leakage signal in magnetic flux leakage type nondestructive testing}
\author{Lintao Zhang$^1$}
\ead{L.Zhang@swansea.ac.uk}
\author{Fawzi Belblidia$^1$}
\author{Ian Cameron$^1$}
\author{Johann Sienz$^1$}
\author{Matthew Boat$^2$}
\author{Neil Pearson$^2$}
\address{$^1$Advanced Sustainable Manufacturing Technologies (ASTUTE) project, College of Engineering, Swansea University, Singleton Park, Swansea SA2 8PP, UK}
\address{$^2$Silverwing (UK) Ltd; Unit 31 Cwmdu Industrial Estate, Swansea, SA5 8JF, UK}
\begin{abstract}
We investigate the influence of the specimen velocity on the magnetic flux leakage with the aim of selecting the optimum sensor locations. Parametric numerical simulations where the specimen velocity was in the range [0.1-20] m$\cdot$s$^{-1}$ were carried out.  As the specimen velocity is increased, the magnetic field varies from being symmetrical to being asymmetric.  For the radial magnetic induction, the peak to peak value moves from the centre of the bridge towards the direction of the specimen movement.  For the axial magnetic induction, the specimen velocity influence is dependent on the sensor location and a signal-velocity independent region was discussed. 
\end{abstract}
\end{frontmatter}
\section{Introduction}
\label{sec:introduction}
Effective nondestructive testing (NDT) can prevent disasters similar to the Buncefield incident \citep{2005b}, resulting in an estimated loss of \pounds 894 million. A great number of NDT methods have been developed. These include: visual, radiographic, ultrasonic, eddy current, thermal infrared, acoustic emission, and magnetic particle testing, \textit{etc.}. In this paper, we focus on the magnetic flux leakage (MFL) method, which originates from the magnetic particle technique. The principle of the MFL method can be understood as follows: when a magnetic field is applied to a ferromagnetic material, the leakage of the magnetic field which is caused by the discontinuity of the geometry can be captured by the sensors, such as Hall probe \textit{etc.}.
The main reason for the leakage is due to the difference, $\sim \mathcal{O}$(10), of the magnetic permeability of the mediums at the interface \citep{2013ysun}. The first application of MFL can be dated back to 1868, the British navy used a compass to examine a magnetised cannon to search for defects \citep{1980gd}. This technique was extensively applied due to the development of magnetisation techniques for examining defects in pipelines, pressure vessels, and wheels, $etc.$ in the 1960s. The defect characteristics, such as shapes, dimensions and locations, can be determined by the leakage signals and a large amount of relevant numerical and experimental research has been carried out. Practically, the nature of such NDT problems is transient rather than static and simply using the results obtained from static simulations to predict transient problems may cause errors, especially under the condition of high velocity. A velocity induced current can be generated by a conducting material moving in a magnetic 
field \citep{1992sniikura} and this phenomenon can alter the distribution of the magnetic field \citep{1997yshin, 1995gk, 2014pwang, 2006yli}. The distortion of the magnetic field points in the specimen movement direction and is independent to the orientation of the magnetizing source. This signal deformation can influence the efficiency of the NDT, especially for determining the defect location. With the aim of compensating for this leakage signal deformation, a scheme was validated against experimental results \citep{2004gspark}. However, the optimum location of the sensor is still not well defined, especially for high velocities. In order to cover an extensive range of parameter in term of specimen velocity, we shall combine a detailed analysis of a pure two-dimensional (2D) geometry with targeted numerical simulations of the full three-dimensional (3D) problem \cite{1997yshin, 2006yli, 1996gk}. Clearly, this is the task which shall be undertaken in this paper.\\
We shall first describe the configuration and numerical setup in section \ref{sec:cas}. In section \ref{sec:dist}, we investigate the distortion of the magnetic field due to the specimen velocity effect. We then discuss the influence of the specimen velocity on the radial magnetic induction $B_y$ in section \ref{sec:by} and axial magnetic induction $B_x$ in section \ref{sec:bx}. General conclusions are summarized in section \ref{sec:concl}. 
\section{Configuration and numerical setup}
\label{sec:cas}
\subsection{Configuration}
\label{sec:cas1}
\begin{figure}[thp!]
\centering
\includegraphics[width=5.5in]{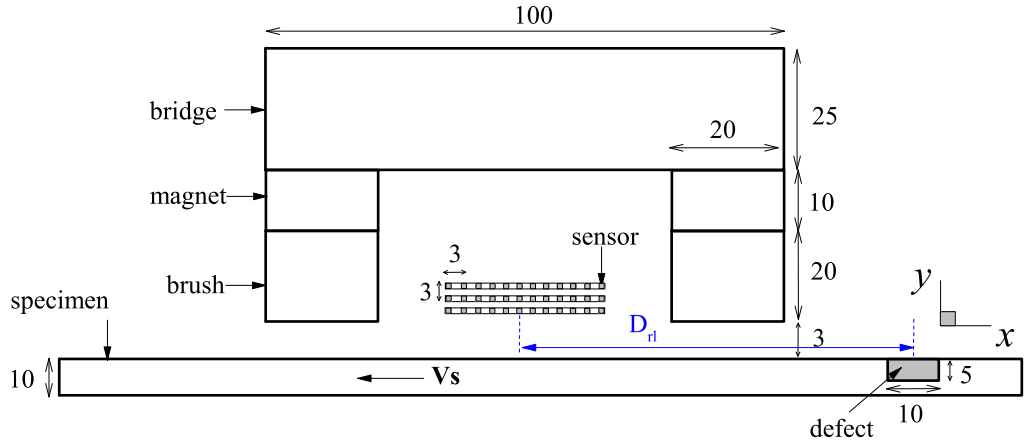}
\vspace{-15mm}
\caption{Schematic representation of numerical simulation model for MFL evaluation system. All dimensions are in mm. The materials of bridge and brush are identical to the specimen. $D_{rl}$, is defined as the relative distance between S$_i^j$ and the centre of the defect. The left edge ($resp.$ right edge) of the defect is defined as front edge ($resp.$ back edge).}
\label{fig:geometry}
\end{figure}

We consider a ferromagnetic specimen (conductivity $\sigma=6.993 \times 10^6$ S$\cdot$m$^{-1}$, having B-H curve) moving with velocity $\textbf{V}_s$ (in m$\cdot$s$^{-1}$ in this paper) under a magnetic flux leakage evaluation system which uses permanent magnets as the magnetising source as shown in Fig.\ref{fig:geometry}.
The direction of specimen movement is taken as the $x$-axis. A near-side rectangular defect is located on the specimen and the reason for the selection of a rectangular shaped defect is because 2D finite element methods (FEM) can provide sufficient information for the sharp-shaped defect characterisation \cite{2006ylij}. The dimensions of the bridge, magnet, brush and the defect are shown in the figure as well. \\
S$_i^j$ denotes the potential sensor locations where $i$ $\in$ [1,11] (along +$x$ axis) and $j$ $\in$ [1,3] (along +$y$ axis), where $i$ and $j$ are the column and row numbers. These potential sensor locations S$_i^j$ are equally spaced 3 mm apart. Furthermore, the relative distance between S$_i^j$ and the centre of the defect is defined as $D_{rl}$. $D_{rl} > 0$ ($resp.$ $D_{rl} < 0$) denotes the centre of the defect is approaching ($resp.$ is departing away from) the sensor.
\subsection{Governing equations and numerical setup}
\label{sec:cas2}
\subsubsection{Governing equations}
We study the 2D magnetic flux leakage problem using Ansoft Maxwell (version 14.0) FEM software.
The governing equations for the 2D transient MFL problem can be expressed as \cite{help}:
\begin{equation}
\label{equ:control_1}
 \nabla \times \frac{1}{\mu} \nabla \times \textbf{A}=\textbf{J}_s-\sigma\frac{\partial \textbf{A}}{\partial t}- \sigma \nabla V+\nabla\times \textbf{H}_c+\sigma \textbf{V}_s\times \nabla \times \textbf{A},
\end{equation}
\begin{equation}
\label{equ:control_2}
 \textbf{B}=\nabla \times \textbf{A},
\end{equation}
where $\mu$, $\textbf{A}$, $\textbf{J}_s$, $\sigma$, $V$, $\textbf{V}_s$ and $\textbf{H}_c$ are the permeability, magnetic vector potential, source current density, electric conductivity,  electric potential, velocity and the coercivity of the permanent magnets, respectively. For the MFL evaluation system investigated in this paper, Equation \ref{equ:control_1} becomes:
\begin{equation}
\label{equ:control_3}
 \nabla \times \frac{1}{\mu} \nabla \times \textbf{A}=-\sigma\frac{\partial \textbf{A}}{\partial t}+\nabla\times \textbf{H}_c+\sigma \textbf{V}_s\times \nabla \times \textbf{A},
\end{equation}
which is solved using FEM with infinite boundary conditions, together with Equation \ref{equ:control_2}. The whole computing domain is discretised into 2D triangular elements. and the first-order implicit Euler method is adopted as the time discretization method. A computational domain percentage, 200\% \cite{help}, is selected.
\subsubsection{Numerical setup}
We first performed a mesh sensitivity analysis to determine the mesh density required for the simulations to give a mesh independent solution. The main characteristics of the meshes tested are provided in Table \ref{123}. $B_y$ and $B_x$ denote the radial and axial magnetic induction.
\begin{table}
\begin{center}
\caption{Main characteristics of the different meshes and errors in the magnetic induction for $\textbf{V}_s$=5 m/s. }\label{123}
\begin{tabular}{llll}
\hline
Meshes&M1&M2&M3\\
\hline
Number of nodes between two brushes&2526&10322&39954\\
Total number of nodes&2.1$\times$10$^4$&8.1$\times$10$^4$&1.1$\times$10$^5$\\
$\epsilon_{B_x}^{4}$&1.3$\times$10$^{-3}$&3.5$\times$10$^{-4}$&-\\
$\epsilon_{B_y}^{4}$&2.2$\times$10$^{-2}$&3.1$\times$10$^{-3}$&-\\
$\epsilon_{B_x}^{6}$&9.6$\times$10$^{-4}$&3.2$\times$10$^{-4}$&-\\
$\epsilon_{B_y}^{6}$&1.3$\times$10$^{-2}$&3.0$\times$10$^{-3}$&-\\
$\epsilon_{B_x}^{8}$&1.2$\times$10$^{-3}$&3.6$\times$10$^{-4}$&-\\
$\epsilon_{B_y}^{8}$&2.3$\times$10$^{-2}$&3.2$\times$10$^{-3}$&-\\
\hline
\end{tabular}
\end{center}
\end{table}
We compute the errors $\epsilon_{Bx}$ (respectively $\epsilon_{By}$) on $B_x$ (respectively $B_y$) relative to the finest mesh M3 at different sensor locations.  $\epsilon_{B_x}^{i}$ and  $\epsilon_{B_y}^{i}$ denote the discrepancies of $B_x$ and $B_y$ between different meshes at the sensor point S$_i^1$: e.g. $\epsilon_{B_x}^{6}=|1-\frac{B_x(M_k)}{B_x(M3)}|$, where $k$=1 and 2. Both these decrease when the total number of mesh elements increases, which shows good convergence. Therefore in this paper, all the simulations are based on mesh M2, which ensures a satisfactory precision at a reasonable computational cost. For M2, the length of element size 0.5 mm for the bridge, magnet, brush and the specimen.\\
A further validation test was carried out to determine the required setup for the transient simulations. This was done by comparing the results obtained from the static solver and the results obtained by using the transient solver with no velocity (\textbf{V}$_s$=0). We compare $B_x$ at different sensor points (S$_i^1$, $i$=1, 3, 5 and 6) at $D_{rl}=0$ and the results show that the discrepancies of $B_x$ are $4.9\times$10$^{-3}$, $3.5\times$10$^{-3}$, $2.5\times$10$^{-3}$ and $2.1\times$10$^{-3}$, respectively.  Therefore, the numerical setup is valid. 
\section{Results}
\label{sec:dist}
\begin{figure}[thp!]
\centering
\includegraphics[width=3.1in]{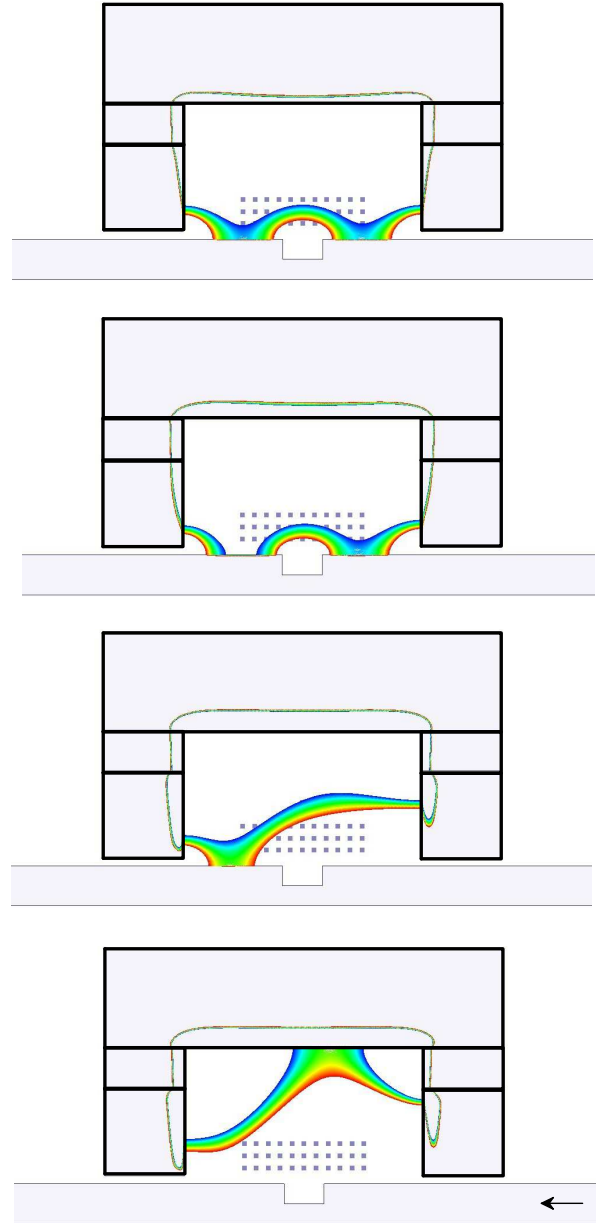}
\caption{Snapshots of distributions of magnetic flux lines when $D_{rl}=0$ at S$_6^1$. From top to bottom: $\textbf{V}_s=0.1, 1, 10$ and 20. The maximum value is -0.0144 (in red) and minimum value is -0.0146 (in blue) Wb/m. The magnetic field distortion occurs as $\textbf{V}_s$ is increased.}
\label{fig:lines1}
\end{figure}

Compared to the static simulations, one of the most important influences of specimen velocity is the distortion of the magnetic field, as shown in Fig.\ref{fig:lines1}. 
The snapshots are the distribution of the magnetic flux lines and are captured at the moment when $D_{rl}=0$ for S$_6^1$ at different specimen velocity ($D_{rl}$ was defined in Fig.\ref{fig:geometry}). For the low values of the specimen velocity, the magnetic field is symmetric to the centre of the defect, as shown in Fig.\ref{fig:lines1} (a). As $\textbf{V}_s$ increases (b - d), the distortion of the magnetic field occurs. The distortion here introduces the asymmetric feature with respect to S$_6^1$. This distortion is mainly caused by the velocity induced current, which was well discussed \cite{1993yks, 1996gk, 1996sm, 2011hk}. In this paper, we mainly focus on the phenomenon which is resulting from this distortion. 
\subsection{Radial magnetic induction $B_y$}
\label{sec:by}
\subsubsection{General feature}
\label{sec:by:1}
\begin{figure}[thp!]
\centering
\includegraphics[width=4in]{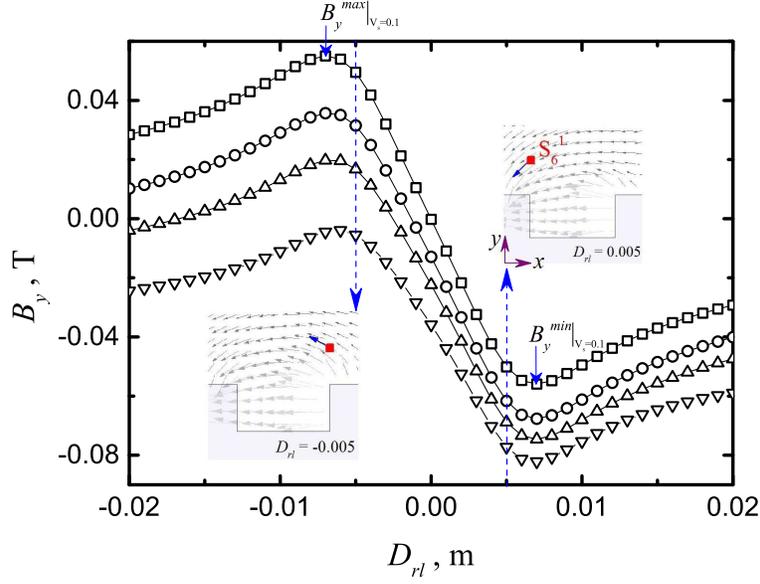}
\caption{Distribution of $B_{y}$ $vs.$ $D_{rl}$ at S$_6^1$ for different $\textbf{V}_s$:  0.1 ($\square$), 5 ($\circ$), 10 ($\bigtriangleup$), 20 ($\bigtriangledown$). The maximum and minimum values are presented when the sensor meets the defect edges. $B_y$ moves downwards as $\textbf{V}_s$ is increased.}
\label{fig:bywithv}
\end{figure}

\begin{figure}[thp!]
\centering
\includegraphics[width=5in]{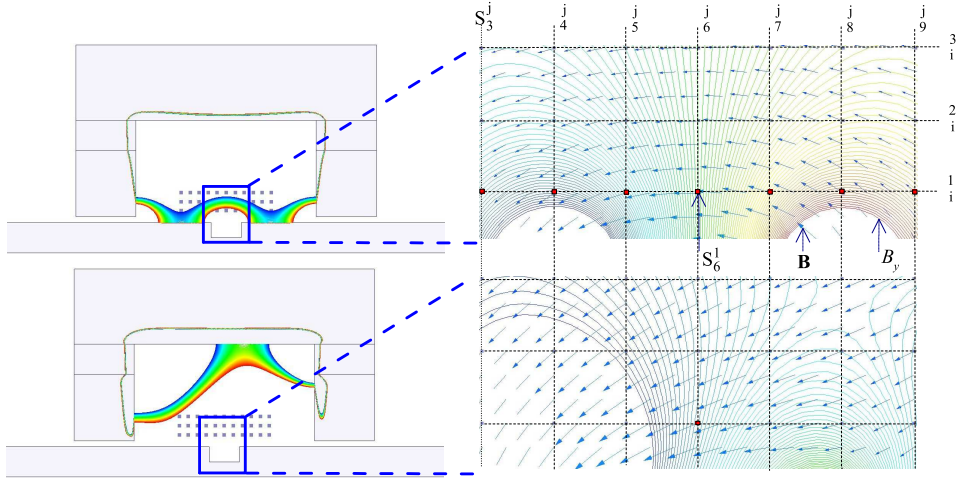}
\caption{Distribution of magnetic induction $\textbf{B}$ and iso-lines of $B_y$ at $D_{rl}$=0 for S$_6^1$ for $\textbf{V}_s$=0.1 ($top$) and 20 ($bottom$). The magnetic induction $\textbf{B}$ rotates anticlockwise as $\textbf{V}_s$ increases.}
\label{fig:bywithv4}
\end{figure}

The distribution of $B_y$ with $D_{rl}$ at S$_6^1$ for different specimen velocity is shown in Fig.\ref{fig:bywithv}.
The result shows that the minimum value of $B_y$ (respectively maximum values of $B_y$) $B_y^{min}$ (respectively $B_y^{max}$) occurs near the front (respectively back) edge of the defect for all values of specimen velocity. This can be understood as follows: in the vicinity of $D_{rl}$=0.005, sensor S$_6^1$ meets the front edge of the defect. The leakage has a trend to return into the specimen due to the high permeability of the specimen. Under this condition, an angle between the $x$-axis and $\textbf{B}$ is presented, which has the effect of enlarging the negative component of $B_y$. When $D_{rl}$=-0.005, S$_6^1$ meets the back edge of the defect. The leakage occurs due to the existence of the forthcoming defect and this results in $B_y$ reaching $B_y^{max}$. Furthermore, distribution of $B_y$ on S$_i^j$, where $i$ $\in$ [1,11] and $j$ $\in$ [1,3], follows the same trend. This indicates that $\textbf{V}_s$ will not influence the effect of the defect edges on $B_y$, compared to the static situation. \\
The distribution of $B_y$ shifts downwards, as specimen velocity is increased. As we discussed before, the distortion of the magnetic field occurs at high value of the specimen velocity. This distortion causes the anticlockwise rotation of the magnetic induction between the two brushes of the MFL evaluation system, as shown in Fig.\ref{fig:bywithv4}. 
Take the magnetic induction at S$_6^1$ as an example: at low values of $\textbf{V}_s$, $\textbf{B}$ is almost parallel to the $x$-axis for the sensor location. The radial component is almost zero. However, at highs value of $\textbf{V}_s$, $\textbf{B}$ rotates anticlockwise and an angle between $\textbf{B}$ and $x$-axis occurs. This angle results in the non-zero magnitude of $B_y$. We find that the variation of $B_y$ obtained at S$_6^1$ is proportional to the specimen velocity (maximum value is 20 m$\cdot$s$^{-1}$).
\subsubsection{Peak to peak value $B_y^p$}
\label{sec:by:2}
\begin{figure}[!htb]
\minipage{0.32\textwidth}
\includegraphics[width=\linewidth]{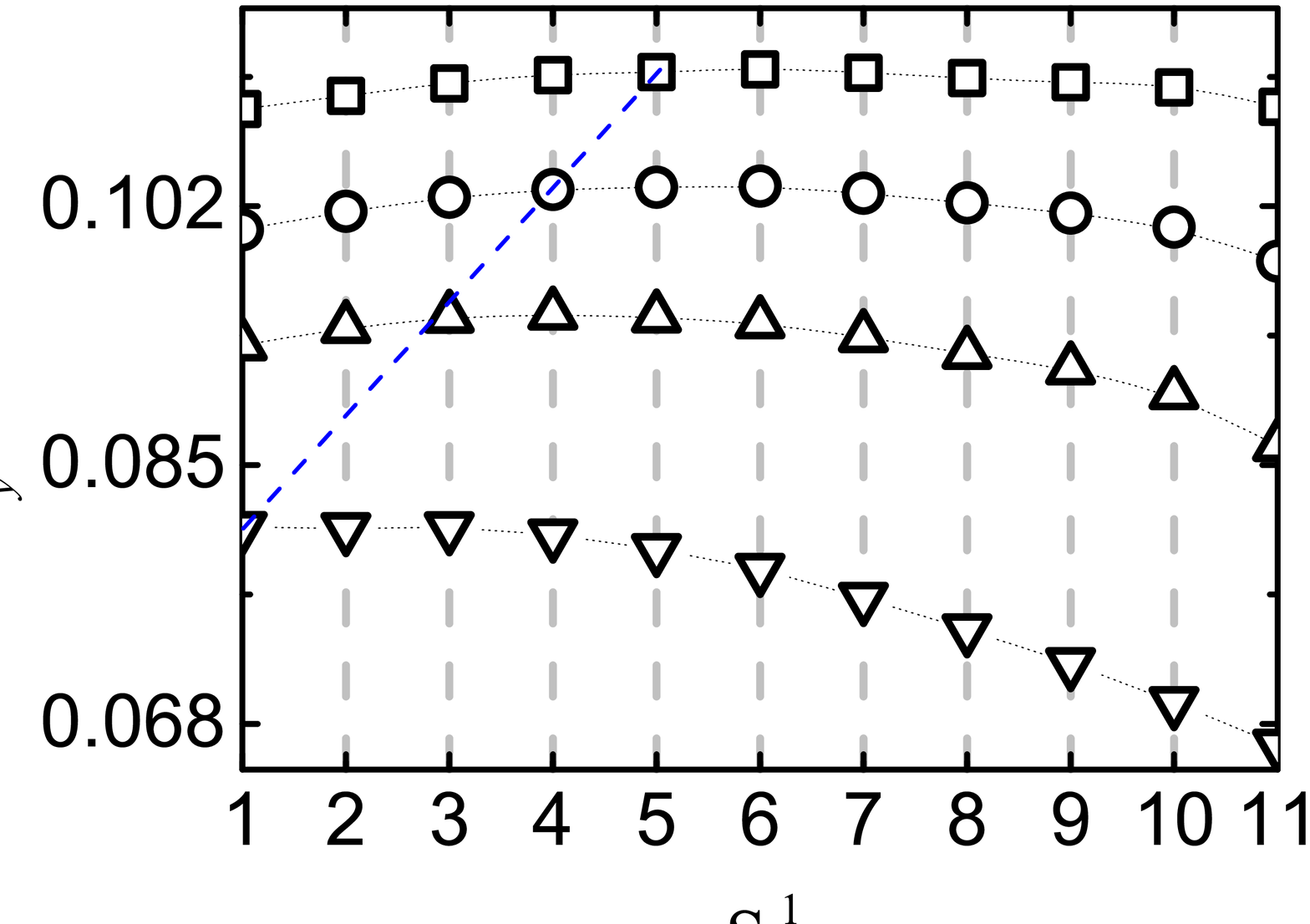}
\endminipage\hfill
\minipage{0.32\textwidth}
\includegraphics[width=\linewidth]{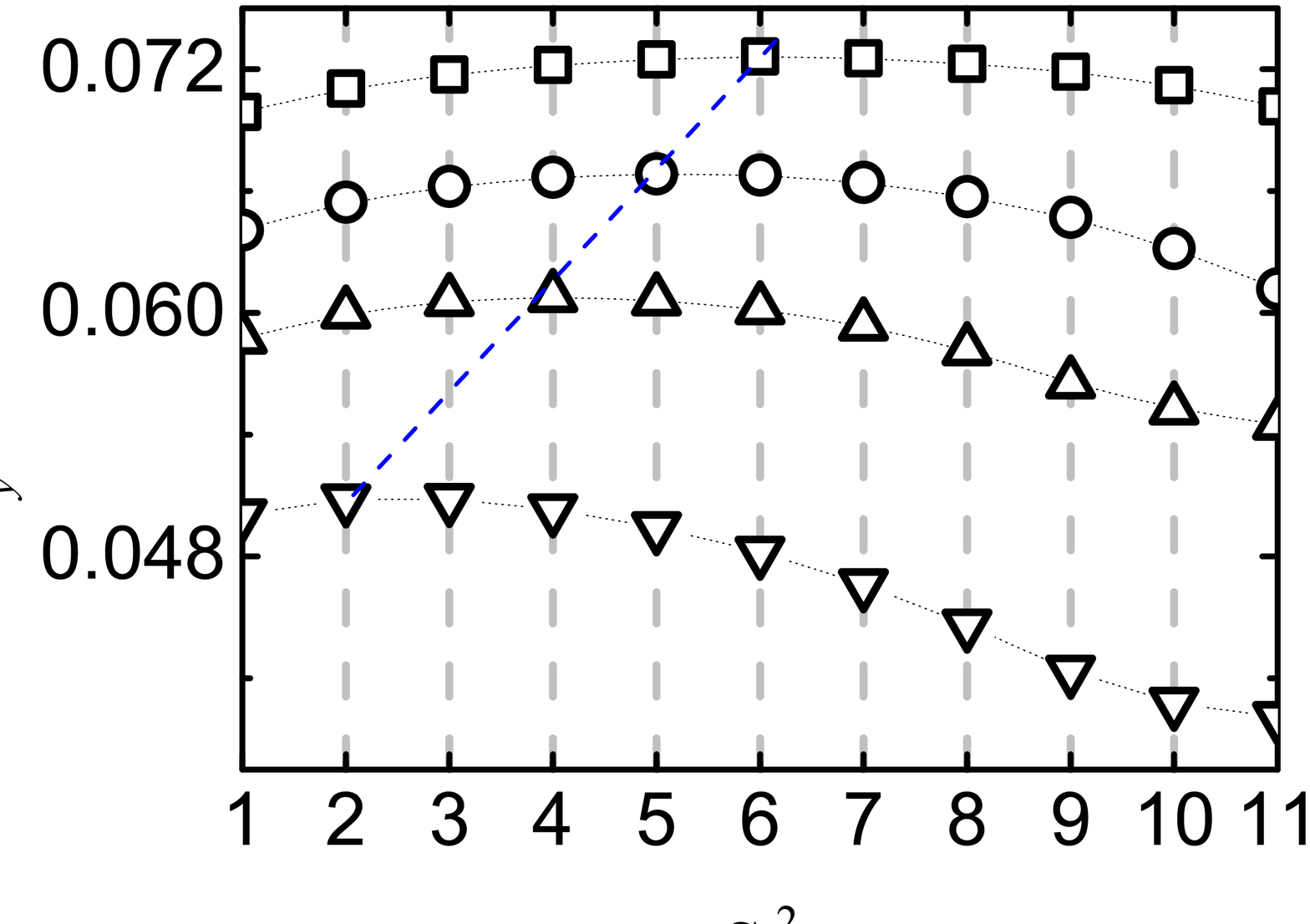}
\endminipage\hfill
\minipage{0.32\textwidth}
\includegraphics[width=\linewidth]{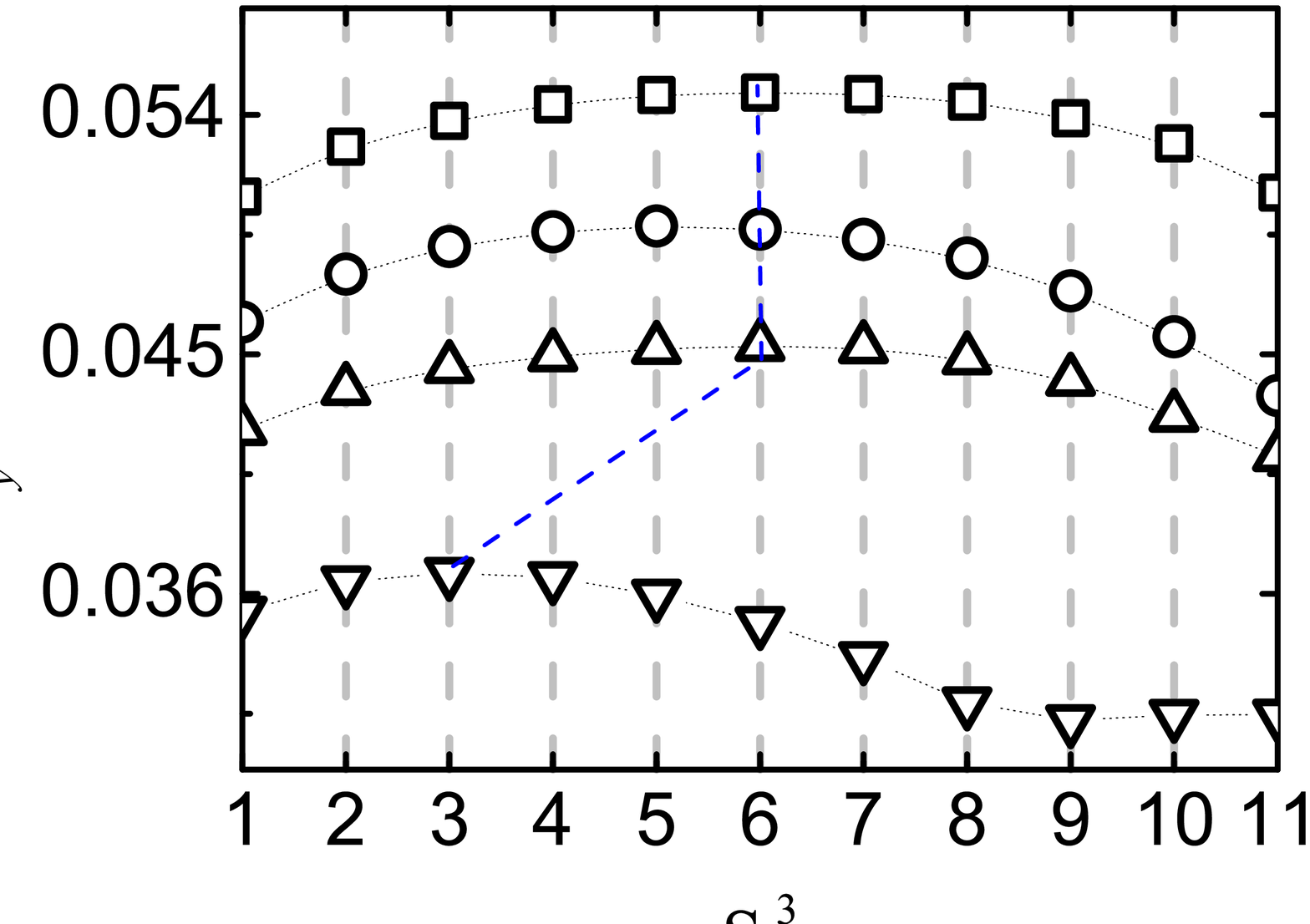}
\endminipage\hfill
\caption{Variations $B_{y}^{p}$ with $S_i^j$ for different $\textbf{V}_s$. $Left$: $j$=1; $middle$: $j$=2; $right$: $j$=3 for $\textbf{V}_s$=0.1 ($\square$), 5 ($\circ$), 10 ($\bigtriangleup$), 20 ($\bigtriangledown$). $B_y^{p}$ moves towards the specimen movement direction as $\textbf{V}_s$ is increased.}
\label{fig:bypp}
\end{figure}

\begin{figure}[thp!]
\centering
\includegraphics[width=5.2in]{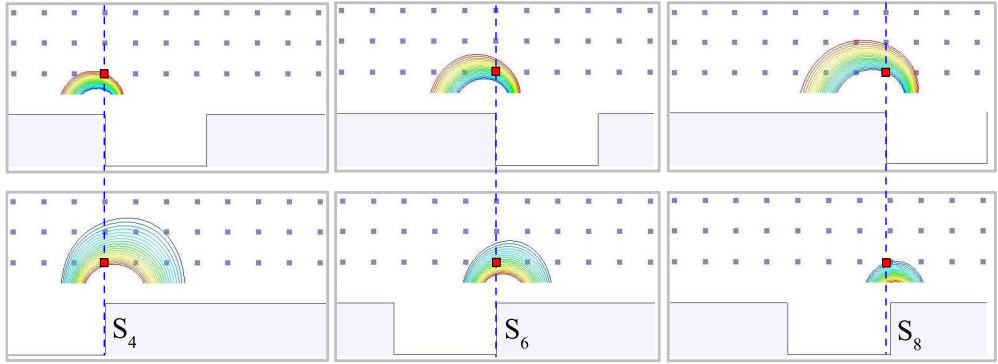}
\caption{Snapshots of $B_y$ distribution when $S_4, S_6$ and $S_8$ meet the front edge ($top$) and back edge ($bottom$). $Top$: the maximum value is -0.058 (in red) and the minimum value is -0.079 (in blue). $Bottom$: the maximum value is 0.03 (in red) and the minimum value is 0.0045 (in blue).}
\label{fig:bymax}
\end{figure}

In practice, the sensitivity of the MFL signal is not only dependent on it's magnitude, but also on the variations of the signal \citep{2002gsp}. We then focus on the variations of the radial leakage signal. The peak to peak value for $B_y$ is defined as follows:
\begin{equation}
B_{y}^{p}=B_y^{max}-B_y^{min}
\end{equation}
The distribution of $B_{y}^
{p}$ at different $S_i^j$ for different $\textbf{V}_s$ is shown in Fig.\ref{fig:bypp}. 
Firstly, the maximum $B_{y}^{p}$ occurs at S$_6^j$ when the specimen velocity is low, e.g. $\textbf{V}_s=$0.1 ($\square$ in the figure), which is located centrally between the two permanent magnets, for all values of $j$. For low specimen velocities, the magnetic field is symmetric about the centre of the defect at $D_{rl}=0$ and $B_y^{max}$ and $B_y^{min}$ are mainly caused by the edges of the defect. $B_y^p$ varies a little at S$_i^j$ and the curve remains relative flat. Secondly, as specimen velocity is increased, the maximum value of $B_{y}^{p}$ occurs further towards the direction of the specimen movement. This phenomenon is mentioned by Shin \citep{1997yshin}, whose findings are based on an MFL system which uses direct current electromagnets as the magnetising, however, the underlying reason for the phenomenon was not well discussed. The phenomenon is mainly due to the distortion of the magnetic field caused by specimen movement. This can be understood as follows. As we mentioned before, $B_y^{min}$ and $B_y^{max}$ occur at the moment when the sensor meets the front edge and back edge, respectively. Fig.\ref{fig:bymax} shows the $B_y$ distribution at the moment when the sensor meets the front and back edge for the sensor S$_4$, S$_6$ and S$_8$ at $\textbf{V}_s$=10.
The results show that from S$_8$ to S$_4$, for $B_y^{min}$, the magnitudes of $B_y$ decreases from 0.079 to 0.058 (26.6\%). However, for  $B_y^{max}$, the magnitudes increases from 0.0045 to 0.03 (566.7\%). This unbalanced increase results in $B_y^p$ moving forwards in the specimen movement direction. For high values of $\textbf{V}_s$, the optimum sensor location is away from the middle of the bridge in the direction of $\textbf{V}_s$.
\subsection{Axial magnetic induction $B_x$}
\label{sec:bx}
\subsubsection{General features}
\label{sec:bx_1}
\begin{figure}[thp!]
\centering
\includegraphics[width=4in]{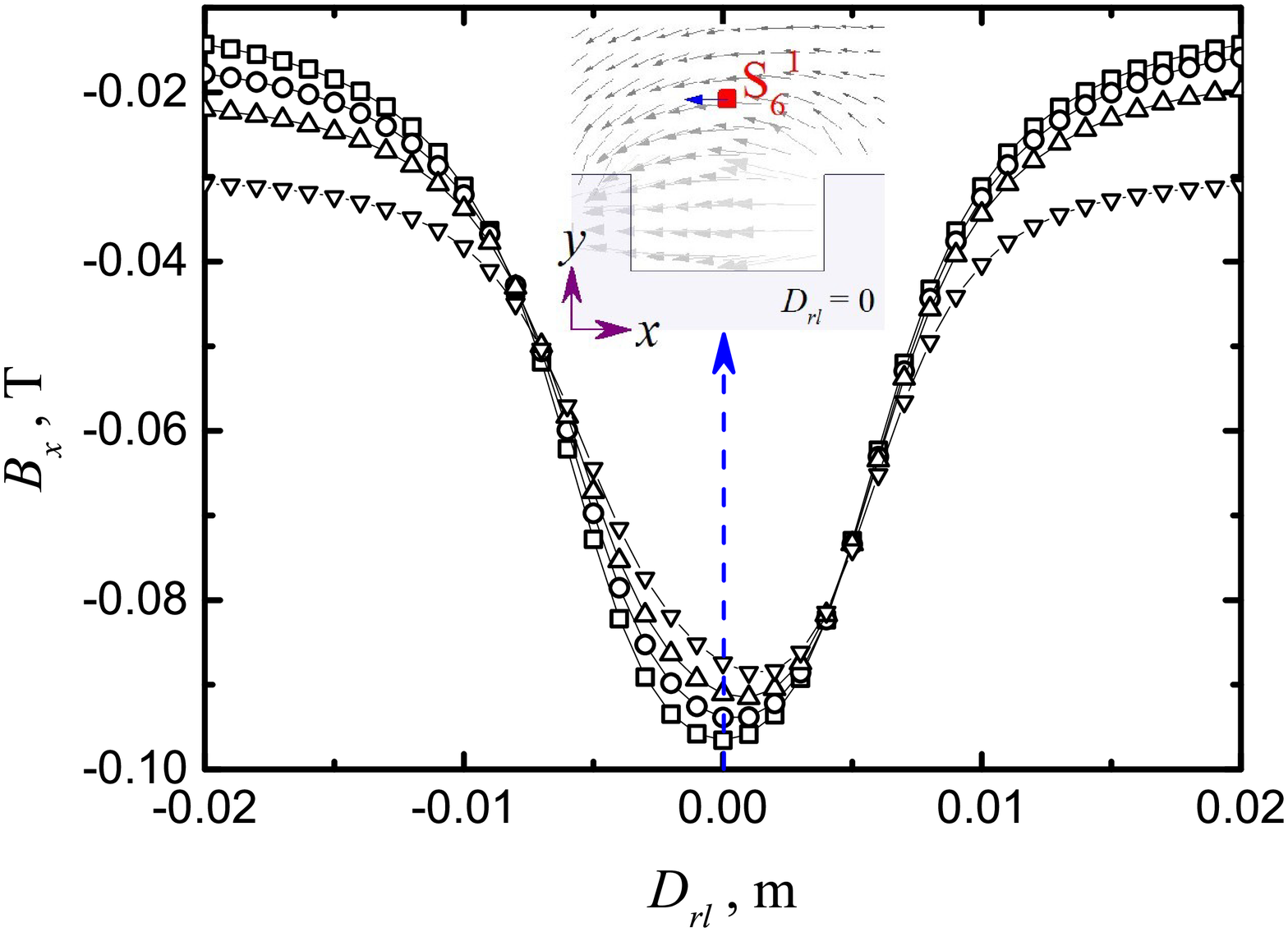}
\caption{Distribution of $B_{x}$ $vs.$ $D_{rl}$ at S$_6^1$ for different $\textbf{V}_s$:  0.1 ($\square$), 5 ($\circ$), 10 ($\bigtriangleup$), 20 ($\bigtriangledown$). For $D_{rl}$=0 mm, $B_x$ decreases as $\textbf{V}_s$ is increased.}
\label{fig:bxwithv}
\end{figure}

The distribution of axial magnetic induction with $D_{rl}$ at the sensor location S$_{6}^1$ for different specimen velocity is presented in Fig.\ref{fig:bxwithv}.
The results show that for all values of $\textbf{V}_s$, the maximum magnitude of $B_x$ occurs in the vicinity of $D_{rl}$=0. This is due to the fact that at that moment $\textbf{B}$ is almost parallel to the $x$-axis. This helps $B_x$ reach its maximum magnitude, as shown in the figure. The variation of $B_x$ at all potential sensor locations S$_i^j$ follows the same trend. As $\textbf{V}_s$ is increased from 0.1 to 20, the maximum magnitude decreases by 11.7\% in the present research.
\subsubsection{$B_x$ on S$_i^j$ for selected $D_{rl}$}
\begin{figure}[thp!]
\minipage{0.48\textwidth}
\centering
\includegraphics[width=2.7in]{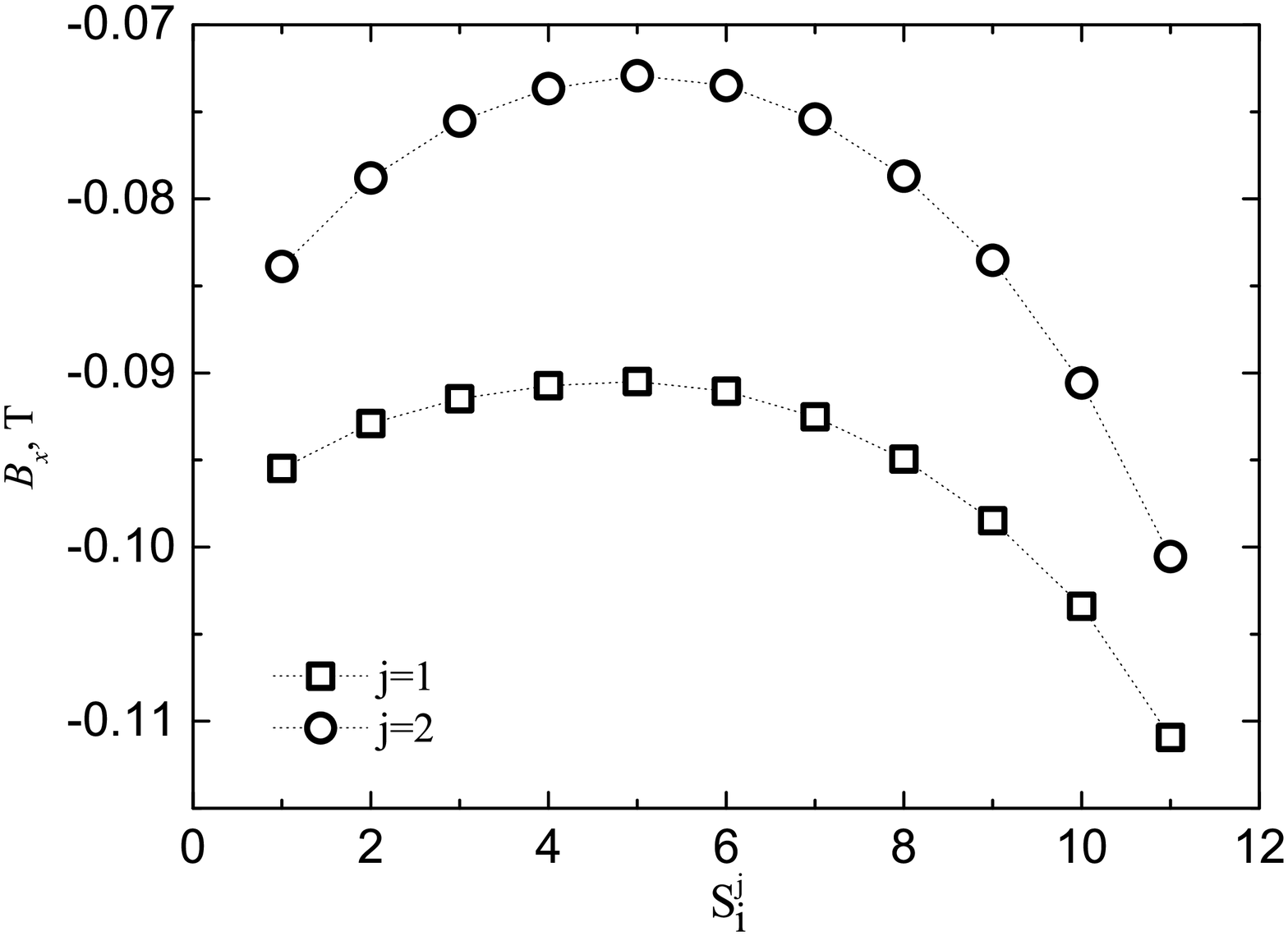}
\endminipage\hfill
\minipage{0.48\textwidth}
\centering
\includegraphics[width=2.6in]{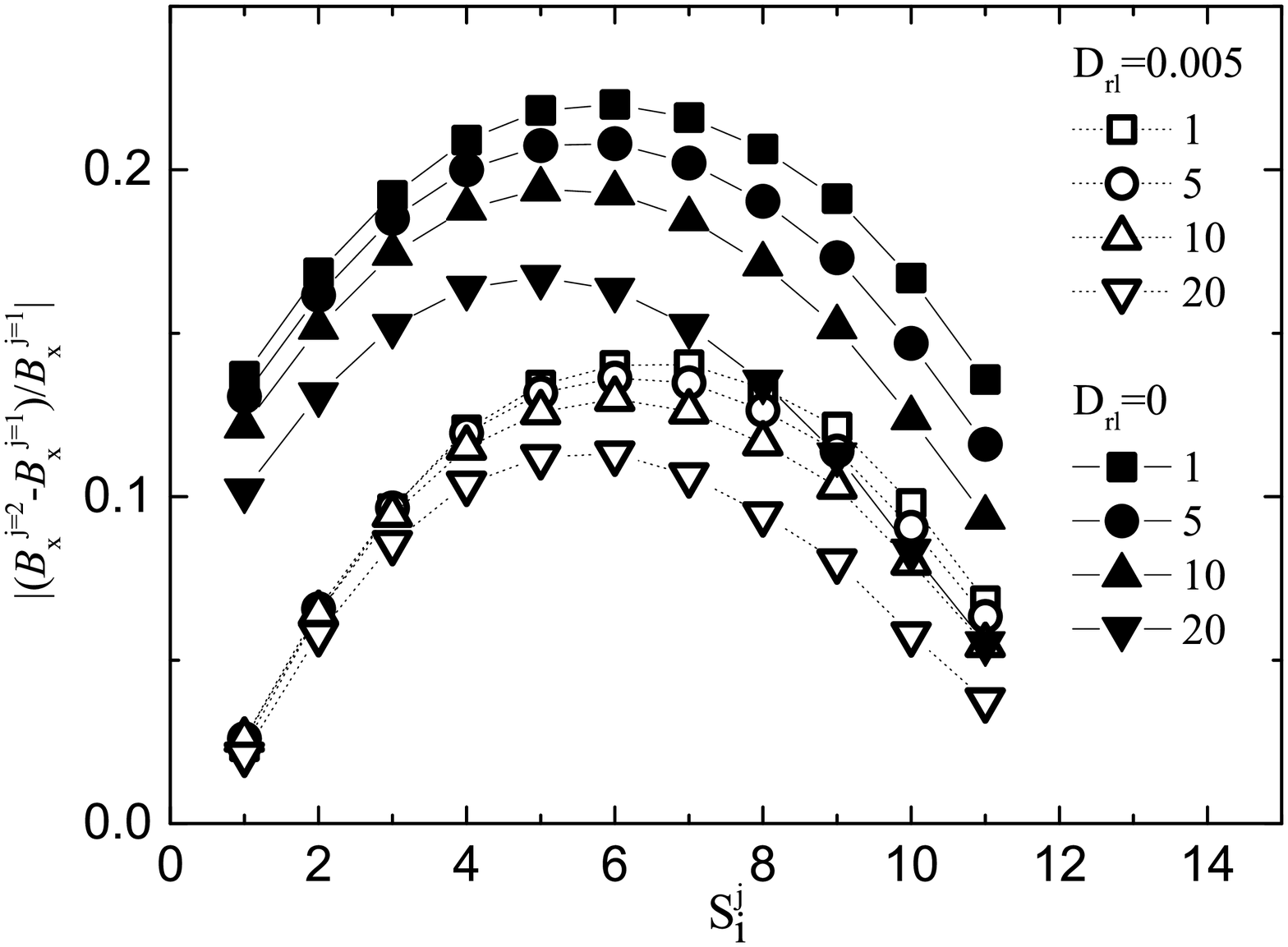}
\endminipage
\caption{\textit{Left}: $B_x$ at different sensor locations for $j$=1 and 2 at $\textbf{V}_s$=10. \textit{Right}:$\abs{\frac{B_x^{j=2}-B_x^{j=1}}{B_x^{j=1}}}$ at different sensor locations for different specimen velocity at $D_{rl}$=0.005 and 0.}
\label{fig:liftoff}
\end{figure}

\begin{figure}[!htb]
\minipage{0.33\textwidth}
  \includegraphics[width=\linewidth]{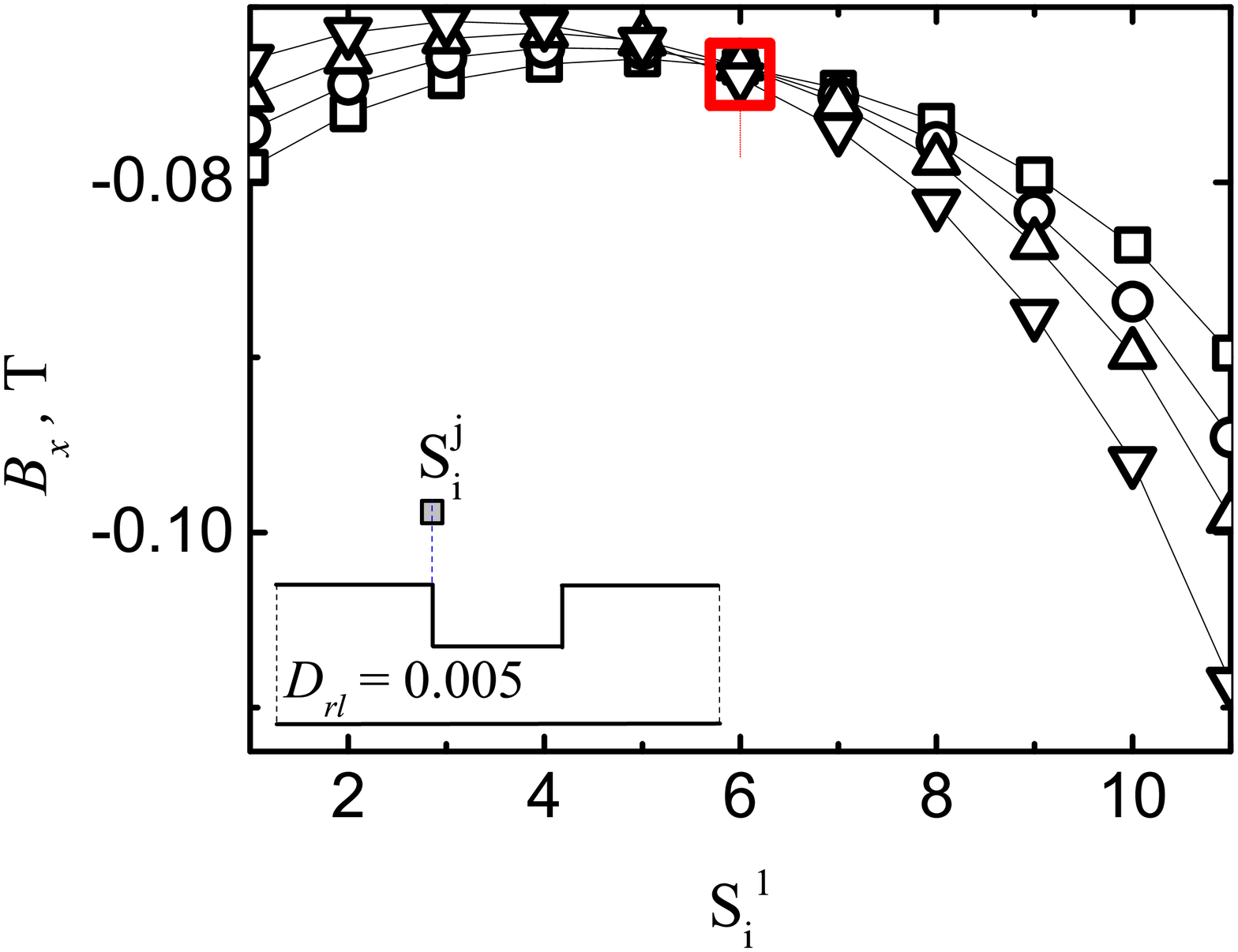}
\endminipage\hfill
\minipage{0.33\textwidth}
  \includegraphics[width=\linewidth]{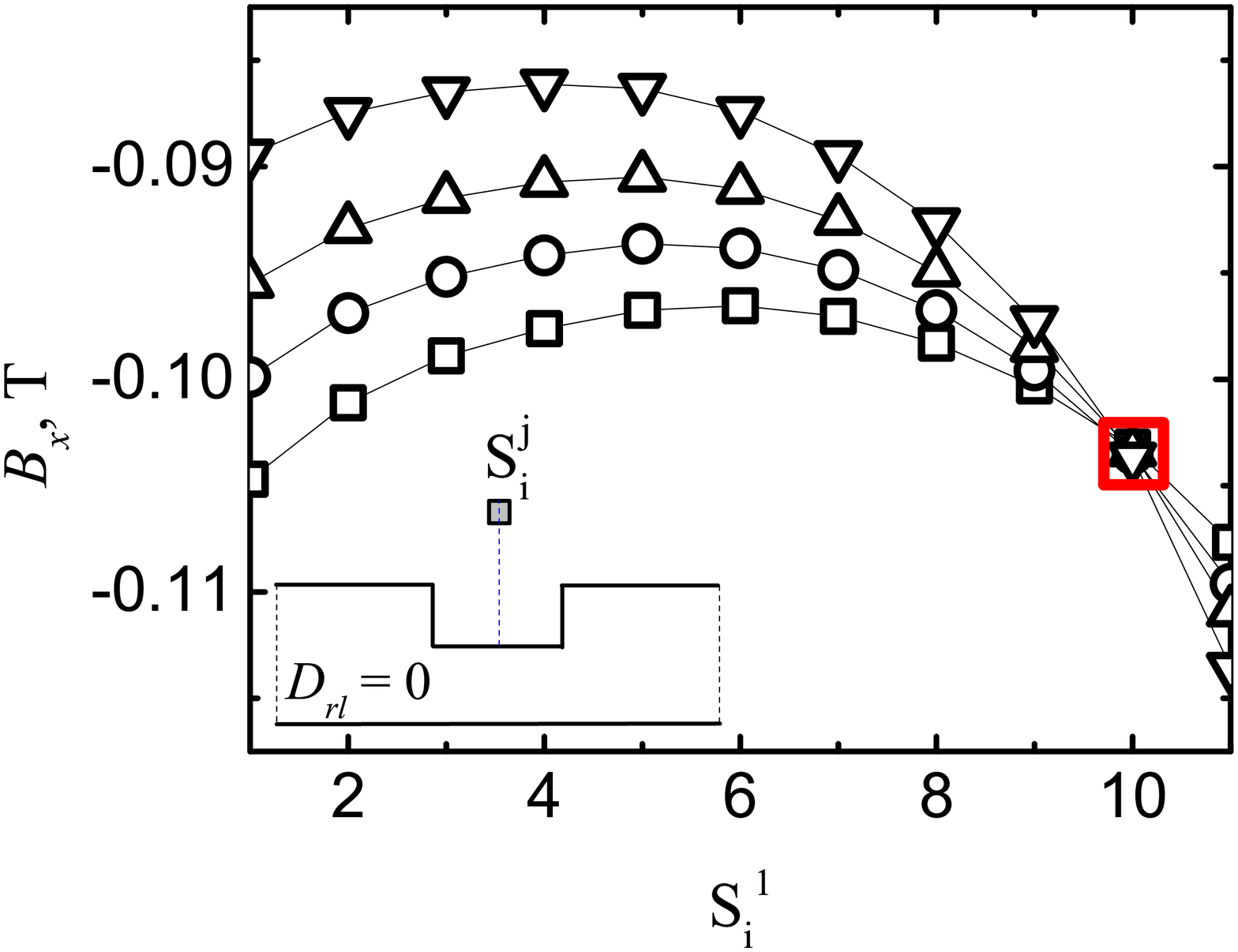}
\endminipage\hfill
\minipage{0.33\textwidth}%
  \includegraphics[width=\linewidth]{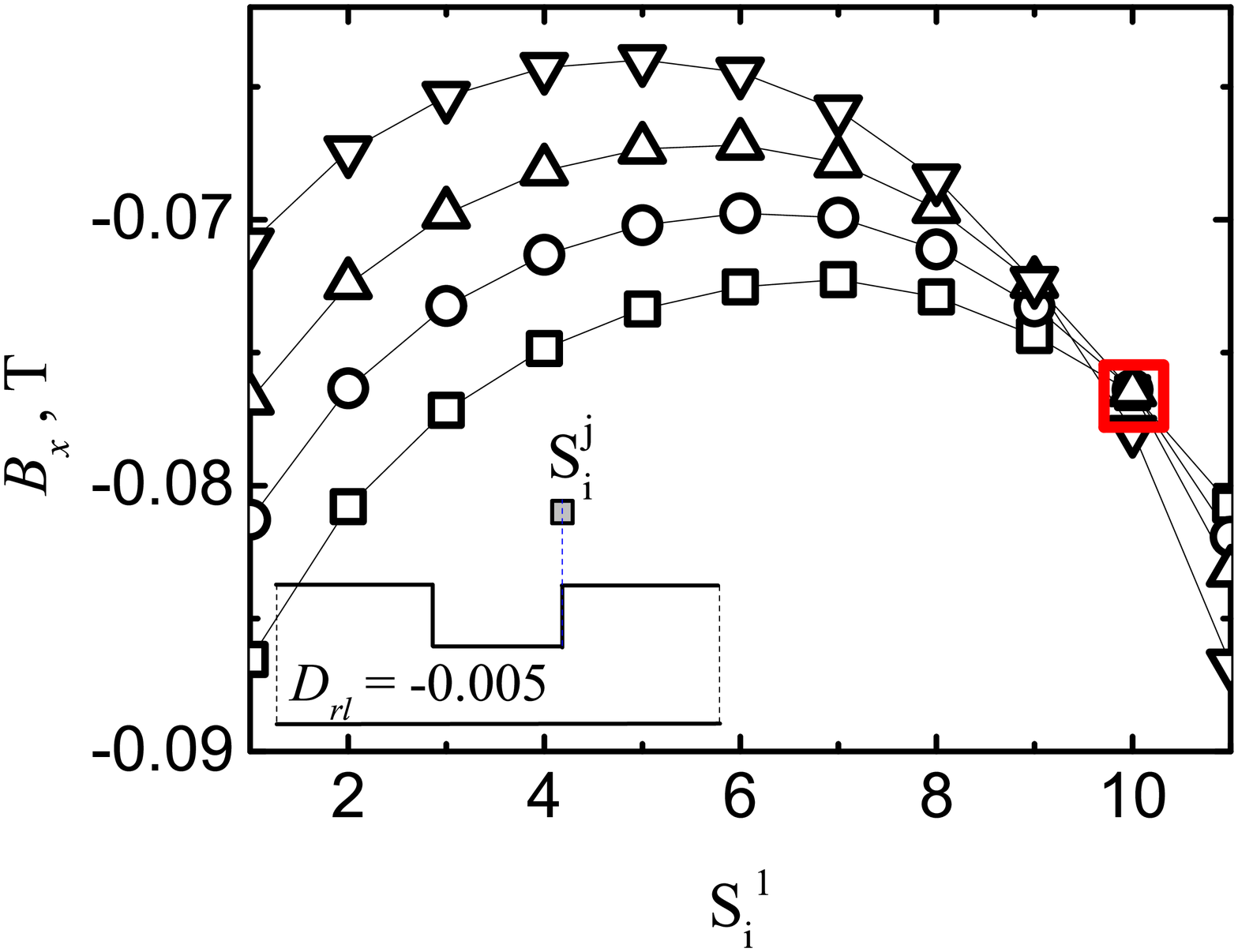}
\endminipage\\
\minipage{0.33\textwidth}
  \includegraphics[width=\linewidth]{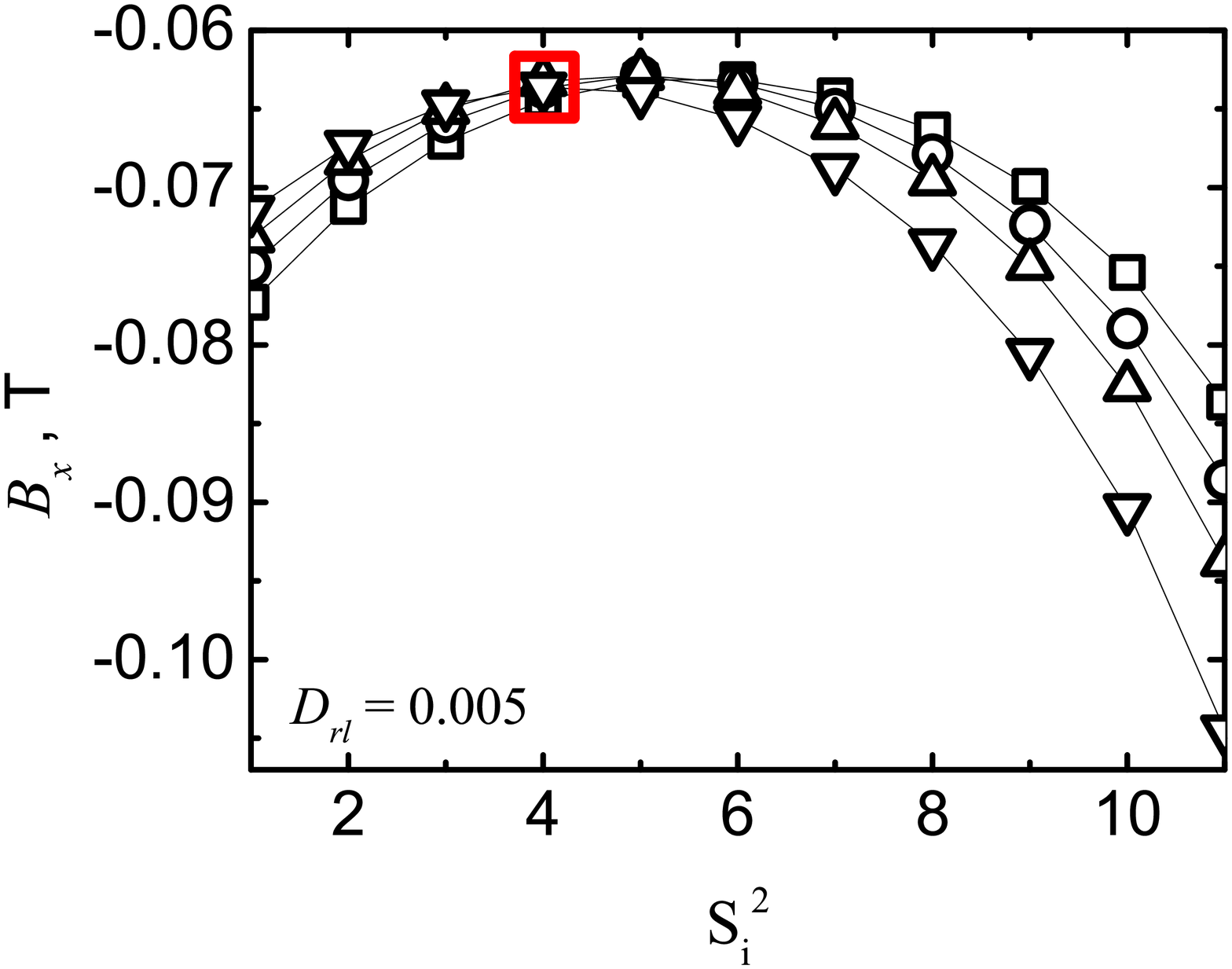}
\endminipage\hfill
\minipage{0.33\textwidth}
  \includegraphics[width=\linewidth]{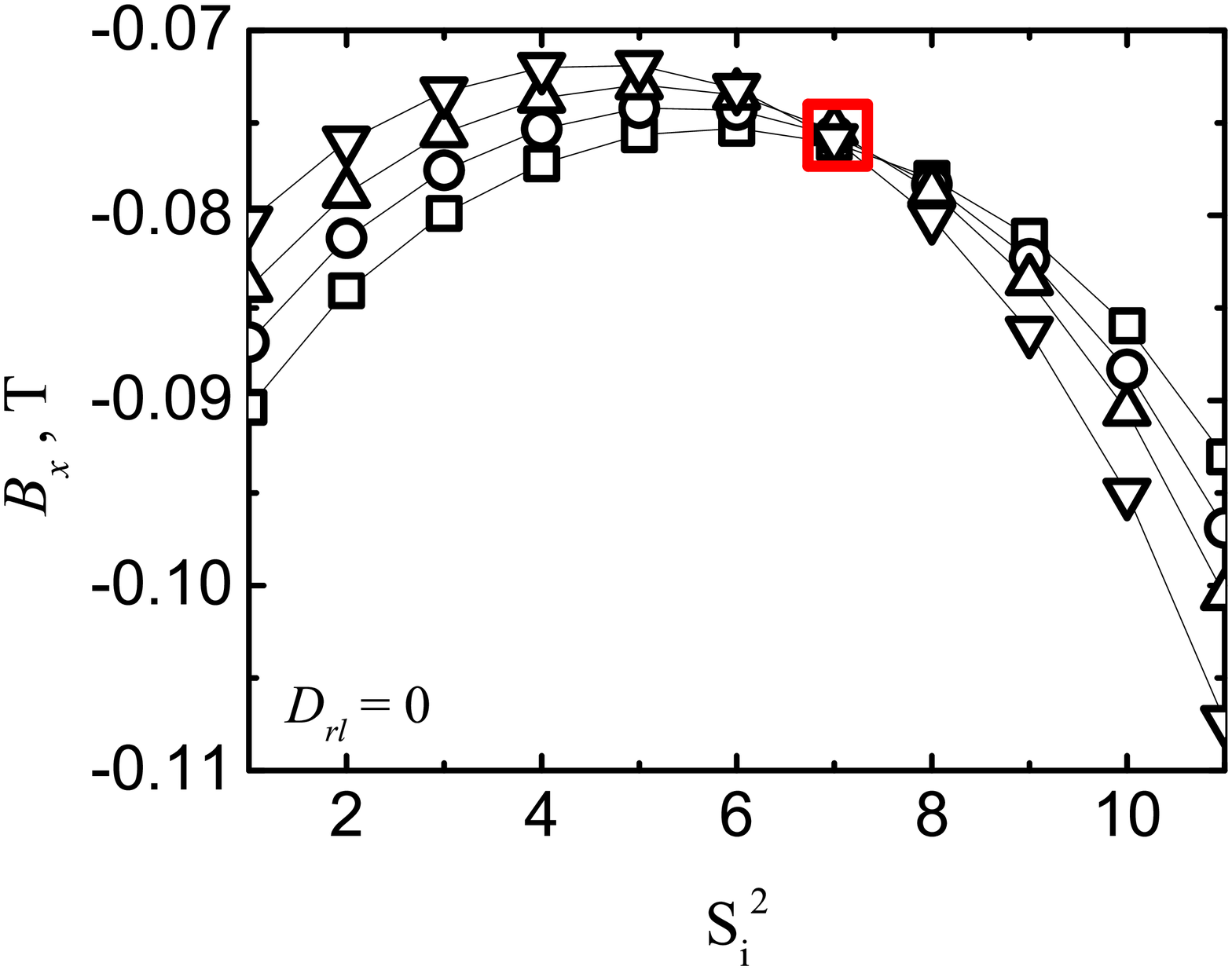}
\endminipage\hfill
\minipage{0.33\textwidth}%
  \includegraphics[width=\linewidth]{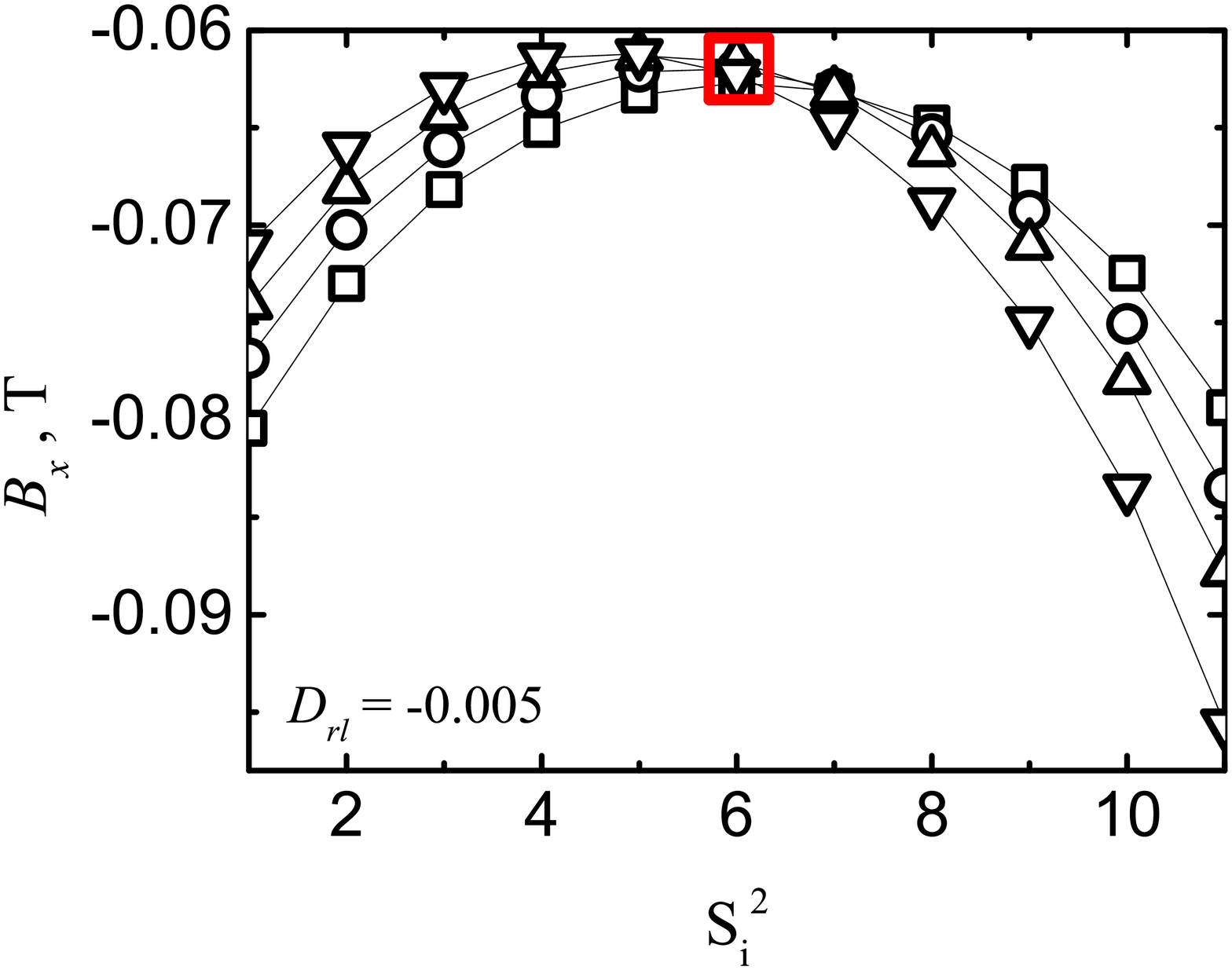}
\endminipage\\
\minipage{0.33\textwidth}
  \includegraphics[width=\linewidth]{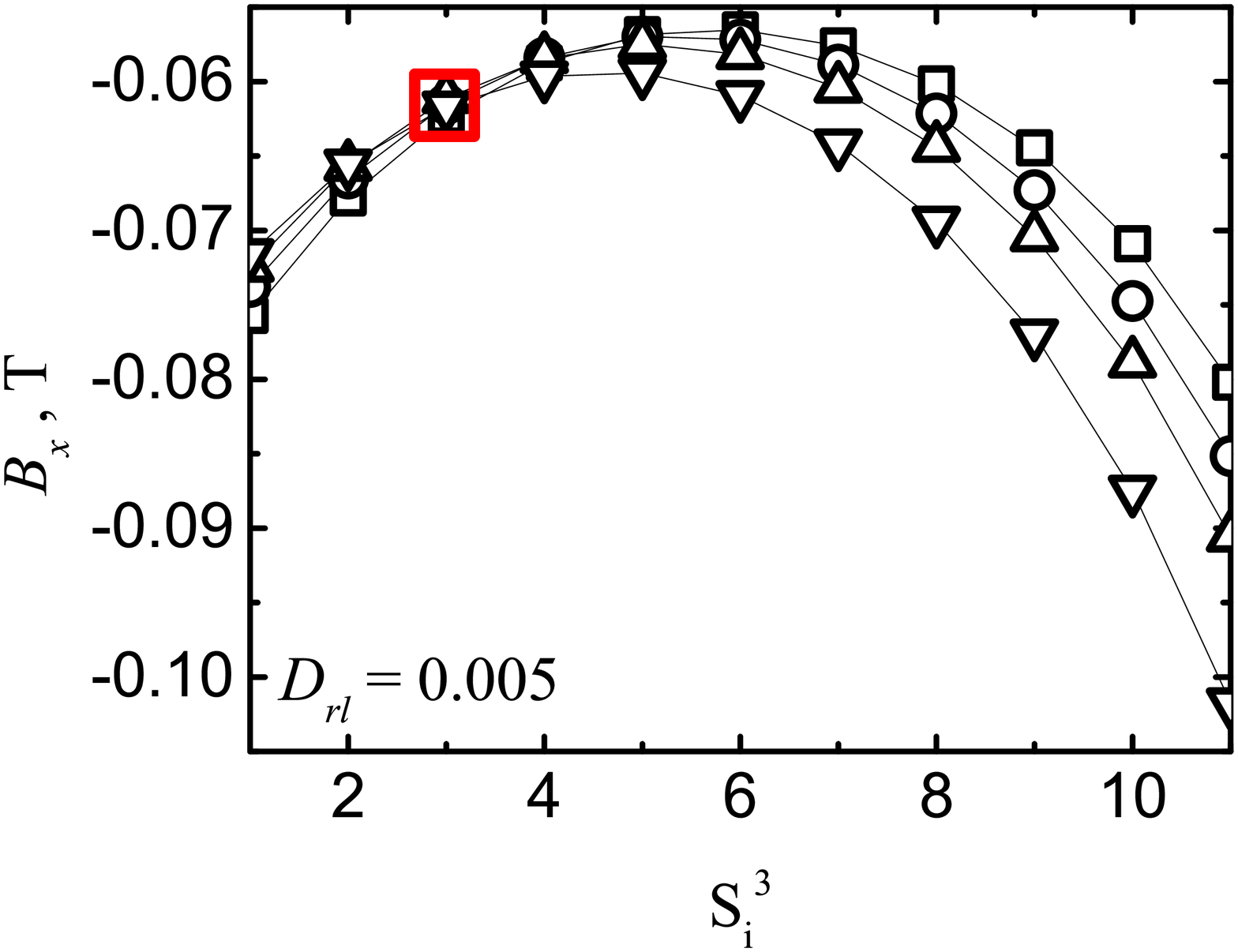}
\endminipage\hfill
\minipage{0.33\textwidth}
  \includegraphics[width=\linewidth]{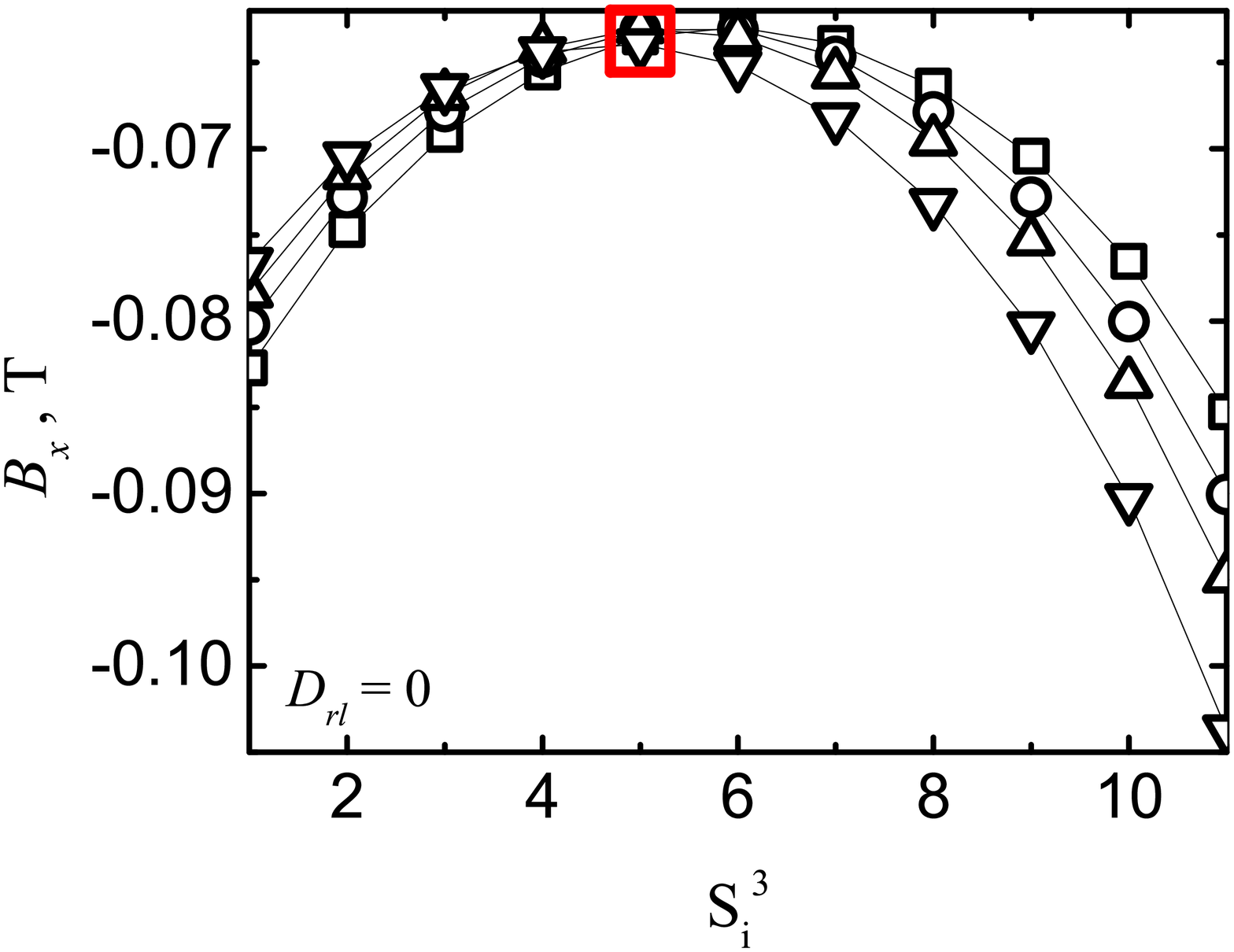}
\endminipage\hfill
\minipage{0.33\textwidth}%
  \includegraphics[width=\linewidth]{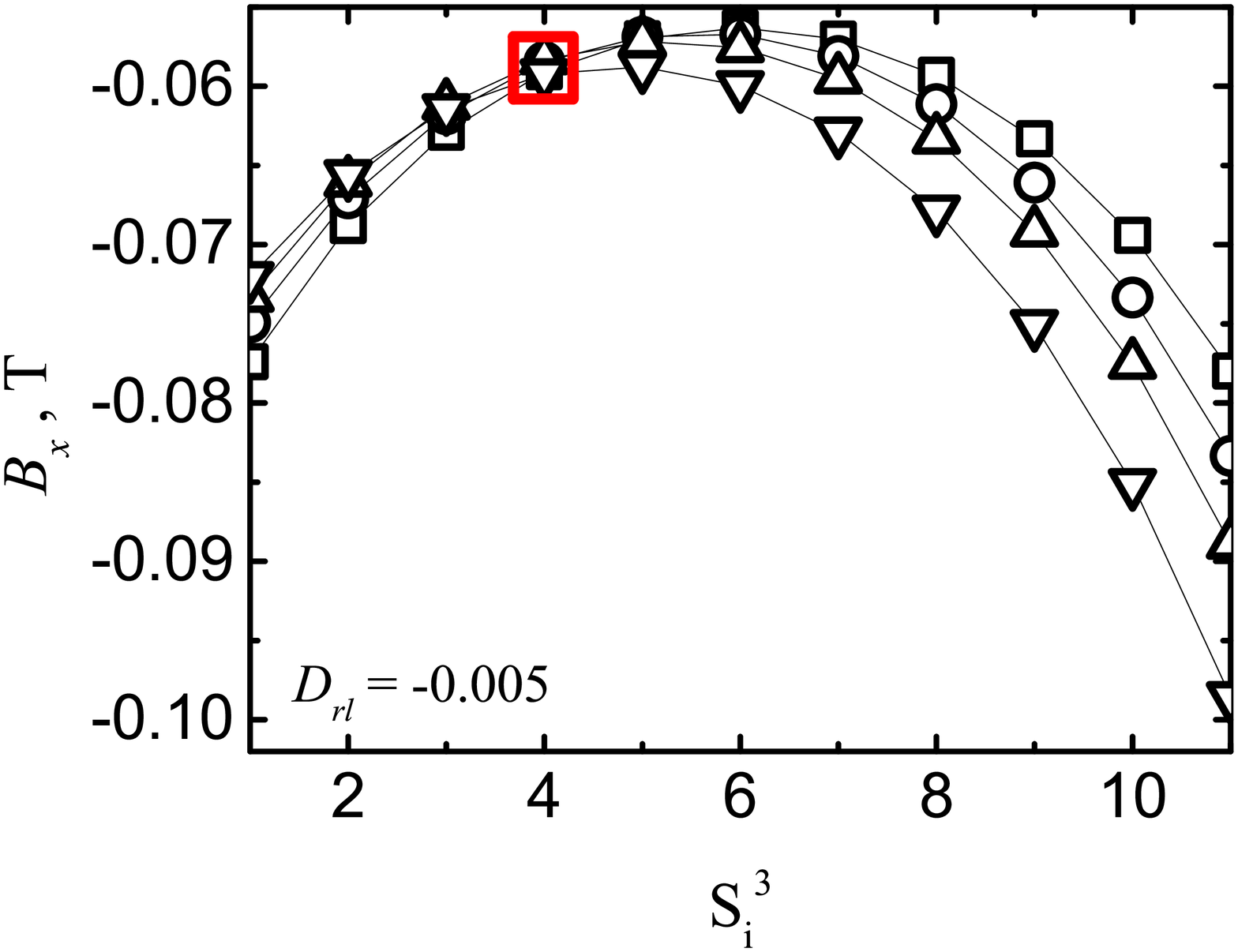}
\endminipage
\caption{Variations of $B_x$ with $S_i^j$ at $D_{rl}$=0.005, 0 and -0.005 for different $\textbf{V}_s$: 1 ($\square$), 5 ($\circ$), 10 ($\bigtriangleup$), and 20 ($\bigtriangledown$). From top to bottom: $j$=1, 2 and 3; from left to right: $D_{rl}$=0.005, 0 and -0.005. Leakage-velocity independent region is highlighted with a red square.}
\label{fig:bxfixeddrl}
\end{figure}

In the section \ref{sec:bx_1}, we focus on the sensor point S$_6^1$. In this section, we turn our attention to the $B_x$ distribution on all the sensor locations.\\
As the lift-off value ($j$) is increased, the magnitudes of $B_x$ decrease (Fig.\ref{fig:liftoff} $left$). This simply reflects the influence of the sensor lift-off values. In order to capture the leakage signal effectively, the sensor should be located near the specimen. However, this phenomenon is weaken as the specimen velocity is increased (Fig.\ref{fig:liftoff} $right$).
\\
The variation of $B_x$ with $S_i^j$ at different specimen velocity for different $D_{rl}$ (0.005, 0 and -0.005) is shown in Fig.\ref{fig:bxfixeddrl}. 
In the figure, from top to bottom, $j$=1, 2 and 3, respectively and from left to right, $D_{rl}$=0.005, 0 and -0.005, respectively. For $\textbf{V}_s$=1 ($\square$ in the figure), the asymmetric (with respect to $B_x$ on S$_6^1$) feature occurs for the selected D$_{rl}$ at $j$=1. The sensor which first meets the defect measures the strongest leakage signal of $B_x$. As $j$ is increased, this feature becomes insignificant. This reflects that increasing $j$ can reduce the influence of $\textbf{V}_s$, however, the strength of the $B_x$ signal is also decreased. As $\textbf{V}_s$ is increased, this asymmetric distribution feature becomes more dominant, even for $j$=3.\\ 
Interestingly, for the geometry we investigated the results show that the influence of $\textbf{V}_s$ on the magnitude of $B_x$ is highly dependent on the sensor location for all values of $D_{rl}$. At $D_{rl}$=0.005 and $j$=1, for the sensor locations $S_i^1$ $i\in$[1,5], the magnitude of $B_x$ decreases as $\textbf{V}_s$ is increased, and conversely, for $S_i^1$ $i\in$[7,11], the magnitude of  $B_x$ increases as $\textbf{V}_s$ is decreased. The velocity has little influence at S$_6^1$ under this condition. Here, we define S$_6^1$ as the leakage-velocity independent region. The existence of this region is mainly due to the differences between $B_x$ gradients for different $\textbf{V}_s$. This location moves in the direction of $\textbf{V}_s$ as $j$ is increased, for all selected $D_{rl}$. In practice, it is helpful to locate the sensor in the vicinity of this region to decrease the influence of $\textbf{V}_s$ from the axial magnetic field point of view.
\section{Conclusions and future work}
\label{sec:concl}
We have conducted a detailed analysis of the influence of specimen velocity on the magnetic flux leakage signal using finite element analysis. The main results can be summarised as follows:
\begin{itemize}
\item deformation of the magnetic field occurs as the specimen velocity is increased, which is due to the velocity induced eddy current effect.
\item For the radial magnetic induction $B_y$, the maximum variation of the leakage signal moves from the centre location between the two brushes towards the specimen movement direction. The reason for this movement is that the leakage that escapes from the specimen increases much faster than the leakage which returns back to the specimen as the specimen velocity is increased. This movement becomes insignificant as the sensor location is further away from the specimen. This indicates that the optimal sensor location is not in the middle of the bridge, but at a location some distance from this point in the direction of the specimen movement, especially for higher speed MFL evaluations.
\item For the axial magnetic induction $B_x$, the influence of the specimen velocity is highly dependent on the sensor location, for the geometry we investigated. A potential sensor location which can capture a constant value of $B_x$ leakage at different specimen velocities is found for the geometry we investigated. This gives us greater flexibility to select suitable sensor locations.
\end{itemize}
Future work will focus on 3D simulations and  experimentally investigating the optimum sensor location for high speed measurement based on peak to peak valued. Furthermore, the leakage-velocity independent region which is based on the axial magnetic induction will be investigated experimentally to confirm the simulation results presented in this paper. 
\section*{Acknowledgement}
The authors would like to acknowledge Advanced Sustainable Manufacturing Technologies (ASTUTE) part-funded by the European Regional Development Fund (ERDF) through the Welsh Government. The authors are grateful to Professor Guiyun Tian, for fruitful discussion with him during the conduct of this work. 


\begin{thebibliography}{100}
\bibitem{2005b} 
Buncefield Major Incident Investigation Board. The Buncefield incident 11 December 2005.
The final report of the Major Incident Investigation Board, 2008.
\bibitem[Y.Sun et al.(2013)]{2013ysun}
Y. Sun and Y. Kang.
Magnetic mechanisms of the magnetic flux leakage nondestructive testing.
\textit{Appl. Phys. Lett.} 2013; 103184104.

\bibitem[G.Dobmann et al. (1980)]{1980gd} G. Dobmann and P. Holler.
Physical analysis methods of magnetic flux leakage. \textit{Academic press}, New York, 1980.

\bibitem[S.Niikura et al. (1992)]{1992sniikura} S. Niikura and A. Kameari.
Analysis of eddy current and force in conductors with motion. 
\textit{IEEE Transactions on Magnetics} 1992; 28(2): p 1450-3. 

\bibitem[Y. Shin (1997)]{1997yshin}
Y. Shin.
Numerical prediction of operating conditions for magnetic flux leakage inspection of moving steel sheets.
\textit{IEEE Transactions on Magnetics} 1997; 33(2): p 2127-30.  

\bibitem[G. Katragadda et al. (1995)]{1995gk} G. Katragadda, Y. S. Sun, W. Load and L. Udipa. Velocity effect and their minimization in MFL inspection of pipelines: a numerical study. New York: Plenum Press Div Plenum Publishing Corp; 1995.

\bibitem[P.Wang et al. (2014)]{2014pwang}
P. Wang, Y. Gao, G. Tian and H. Wang.
Velocity effect analysis of dynamic magnetization in high speed magnetic flux leakage inspection. \textit{NDT\&E International} 2014; 64: p 7-12.  

\bibitem[Y.Li et al. (2006)]{2006yli}
Y. Li, G. Y. Tian and S. Ward.
Numerical simulation on magnetic flux leakage evaluation at high speed.
\textit{NDT\&E International} 2006; 39: p 367-73.

\bibitem[G.S.Park et al. (2004)]{2004gspark}
G. S. Park and S. H. Park.
Analysis of the velocity-induced eddy current in MFL type NDT.
\textit{IEEE Transactions on Magnetics} 2004; 40(2): p 663-6.

\bibitem[G. Katragadda et al. (1996)]{1996gk}
G. Katragadda, W. Lord, Y. S. Sun, S. Udpa and L. Udpa. 
Alternative magnetic flux leakage modalities for pipeline inspection.
\textit{IEEE Transactions on Magnetics} 1996; 32(3): p 1581-4.

\bibitem[Y.Li et al. (2006)]{2006ylij}
G. S. Park and E. S. Park.
Experiment and simulation study of 3D magnetic field sensing for defect characterisation.
\textit{Proceedings of the 12$^{th}$ Chinese Automation \& computing Society conference in the UK}, Loughborough, England 2006.

\bibitem[Help (v14.0)]{help} ANSOFT Corporation. Maxwell V14.0 Manual; 2010.

\bibitem[Y. S. Shin et al. (1993)]{1993yks}
Y. K. Shin and W. Lord. 
Numerical modelling of moving prove effects for electromagnetic nondestructive evaluation. \textit{IEEE Transactions on Magnetics} 1993; 29(2): p 1865-8


\bibitem[S. Mandayam et al. (1996)]{1996sm}
S. Mandayan, L. Udpa, S. S. Udpa and W. Lord. Invariance transformations for magnetic flux leakage signals.
\textit{IEEE Transactions on Magnetics} 1996; 32(3): p 1576-80.

\bibitem[H. Kikuchi et al. (2011)]{2011hk}
H. Kikuchi, K. Sato, I. Shimizu, Y. Kamada and S. Kobayashi. 
Feasibility study of application of MFL to monitoring of wall thinning under reinforcing plates in nuclear power plants. 
\textit{IEEE Transactions on Magnetics} 2011; 47(10): p 3963-6.

\bibitem[G. S. Park et al. (2002)]{2002gsp}
G. S. Park and E. S. Park.
Improvement of the sensor system in magnetic flux leakage-type nondestructive testing.
\textit{IEEE Transactions on Magnetics} 2002; 38(2): p 1277-80.



\end{thebibliography}
\end{document}